\documentclass{aa}
\usepackage{rotating}
\def\deg{^\circ} 

\begin{document}
\thesaurus{6(08.22.2, 08.15.1,08.09.2: FG Vir, 08.09.2: HD 106384, 08.20.4)}
\title{The $\delta$ Scuti star FG Vir.
IV. Mode identifications and pulsation modelling}

\author
{M.~Breger\inst{1} \and A.~A.~Pamyatnykh\inst{1,2,3}
\and H.~Pikall\inst{1} \and R.~Garrido\inst{4}}

\offprints{M.~Breger}

\institute{Astronomisches Institut der Universit\"at Wien,
T\"urkenschanzstr. 17,
A--1180 Wien, Austria\\INTERNET: breger@astro.univie.ac.at
\and
Copernicus Astronomical Center, Bartycka 18, 00-716 Warsaw, Poland
\and
Institute of Astronomy, Russian Academy of Sciences, Pyatnitskaya Str. 48,
109017 Moscow, Russia
\and
Instituto de Astrofisica de Andalucia, CSIC, Apdo. 3004, E-18080 Granada,  
Spain}

\date{Received date; accepted date}

\maketitle
\markboth{M. Breger et al.: FG Vir mode identifications}{M. Breger et al.:
FG Vir mode identifications}

\begin{abstract}

This paper examines the mode identification and presents pulsation models for
FG~Vir, for which 24 frequencies have been detected.
Histograms of the frequency spacings show peaks which are identified with
adjacent radial orders and rotational splitting.

Pulsational $\ell$ values are deduced for eight modes by comparing
the observed photometric phase lags between $v$ and $y$
variations with calculated values. The dominant pulsation
mode at 12.72~c/d can be identified with $\ell = 1$, while the
12.15~c/d mode is the radial fundamental. These results are in agreement with
identifications published by Viskum et al. (1998).

Based on the observational mode identifications and the Hipparcos
distance, new models were computed with the constraint that the
mode at 12.15 c/d is the radial fundamental mode.
It is shown that with standard opacities, models in the appropriate
$T_{\rm eff}$, $\log L$ and $\log g$ ranges 
cannot reproduce the identification
in the literature of 23.40 c/d as the third radial overtone. However,
we show that observationally an $\ell$ = 1 (rather than
radial) identification is equally probable.

A large number of pulsation models were computed for FG~Vir.
A comparison between the observed frequencies and mode identifications
and pulsation models leads to a mean density of 
$\bar{\rho}/\bar{\rho_{\odot}} = 0.156 \pm 0.002$ depending on the
opacity and chemical composition choice and on the possible
overshooting from the convective core. The models also correctly
predict the observed region of instability between 9 and 34~c/d.

The effect of rotational coupling on the pulsation frequencies is
estimated.

\end{abstract}

% *********************************************************************
\section{Introduction}

FG~Vir (=HD 106384) is a $\delta$~Scuti variable near the end of
its main-sequence evolution. 435 hours of photometric
measurements by the Delta Scuti Network determined 24 statistically
significant frequencies  from 9.20 to 34.12~c/d (106 to 395~$\mu$Hz).
Details of this campaign as well as references to earlier measurements
and results can be found in Breger et al. (1998). The large number of
detected pulsation modes makes this star an excellent candidate for
asteroseismological investigations. This requires the identifications of
the observed pulsation frequencies with specific pulsation modes.
While the problem is rather complex, considerable progress has been
achieved, as shown by Breger et al. (1995), Guzik et al. (1998) and 
Viskum et al. (1998). 

\begin{table*}
\begin{center}
\caption{Pulsation frequencies of FG Vir}
\begin{tabular}{lcccc}
\hline
\noalign{\smallskip}
\multicolumn{3}{c}{Frequency}& 1995 V amplitude & $Q$ value \\
& c/d & $\mu$Hz& mmag & days\\
\noalign{\smallskip}
\hline
\noalign{\smallskip}
\multicolumn{3}{l}{Statistically significant frequencies}\\
\noalign{\smallskip}
$f_1$ & 12.716 & 147.2 & 21.1 & .0323\\
$f_2$ & 24.228 & 280.4 & 4.5 & .0170\\
$f_3$ & 23.403 & 270.9 & 4.1 & .0176\\
$f_4$ & 21.052 & 243.7 & 3.7 & .0195\\
$f_5$ & 19.868 & 230.0 & 3.5 & .0207\\
$f_6$ & 12.154 & 140.7 & 3.5 & .0338\\
$f_7$ & 9.656 & 111.8 & 3.4 & .0426\\
$f_8$ & 9.199 & 106.5 & 3.1 & .0447\\
$f_9$ & 19.228 & 222.5 & 1.5 & .0214\\
$f_{10}$ & 20.288 & 234.8 & 1.3 & .0203\\
$f_{11}$ & 24.200 & 280.1 & 1.3 & .0170\\
$f_{12}$ & 16.074 & 186.0 & 1.0 &  .0256\\
$f_{13}$ & 34.119 & 394.9 & 1.0 &  .0121\\
$f_{14}$ & 21.232 & 245.7 & 1.0 & .0194\\
$f_{15}$ & 11.110 & 128.6 & 0.9 & .0370 \\
$f_{16}$ = 2$f_1$& 25.432 & 294.4 & 0.9 & .0162\\
$f_{17}$ & 33.056 & 382.6 & 0.6 & .0124 \\
$f_{18}$ & 21.551 & 249.4 & 0.8 & .0191 \\
$f_{19}$ & 28.140 & 325.7 & 0.6 & .0146\\
$f_{20}$ & 11.195 & 129.6 & 0.7 & .0367 \\
$f_{21}$ & 24.354 & 281.9 & 0.6 & .0169\\
$f_{22}$ & 11.870 & 137.4 & 0.4 & .0346 \\
$f_{23}$ = $f_1+f_7$ & 22.372 & 258.9 & 0.5 & .0184\\
$f_{24}$ = $f_3-f_1$ & 10.687 & 123.7 & 0.5 & .0385\\
\noalign{\smallskip}
\multicolumn{3}{l}{Probable frequencies}\\
\noalign{\smallskip}
$f_{25}$ & 25.37 & 293.7 & 0.4 & .0162\\
$f_{26}$ & 25.18 & 291.4 & 0.4 & .0163\\
$f_{27}$ & 29.50 & 341.4 & 0.4 & .0139\\
$f_{28}$ & 18.16 & 210.2 & 0.4 & .0226\\
$f_{29}$ & 19.65 & 227.4 & 0.4 & .0209\\
$f_{30}$ & 31.92 & 369.4 & 0.4 & .0129\\
$f_{31}$ & 20.83 & 241.1 & 0.4 & .0197\\
$f_{32}$ & 12.79 & 148.1 & 0.4 & .0322\\
\noalign{\smallskip}
\hline
\end{tabular}\newline
\end{center}
\end{table*}

The pulsation mode indentification from observed frequencies
requires accurate determinations of the basic
parameters of the star. From the available $uvby\beta$ photometry,
Mantegazza et al. (1994) derived $T_{\rm eff} = 7500\, $K and
$\log \, g = 3.95$. A correction for a misprint in the literature leads to a
correction of $\log \, g$ to 3.9. We can now improve these
values further by including the accurate Hipparcos parallax
which predicts $M_{\rm V} = 1.95 \pm 0.13\, $mag.
This is slightly fainter than the value of $1.71 \pm 0.25$ mag
predicted by $uvby\beta$ photometry. This leads to a corresponding shift
in $\log \, g$ to 4.00.
These values are in exact agreement with those derived by Viskum et al. (1998).
We estimate the uncertainties to be $\pm 100\, $K in temperature
and $\pm 0.1$ in $\log \, g$.

The values of the pulsation constants $Q$ can be estimated from the
following empiric equation:
\[\log Q_{i} = -6.456 + \log P_{i} + 0.5 \log g + 0.1\,M_{\rm bol} +
\log T_{\rm eff}.\]
\noindent
The constant,
--6.456, in the above formula is based on solar values
of $ M_{\rm bol} = 4.75\, $mag, $B.C. = - 0.08\, $mag,
$T_{\rm eff} = 5770\, $K and $\log \, g = 4.44$.
If the $Q$ values are calculated from $uvby\beta$ photometry,
the observational uncertainties in observing these parameters
lead to an uncertainty in $Q$ of about 18\%.

The corresponding $Q$ values are shown in Table~1.

\begin{figure*}
\centering
\includegraphics*[width=178mm]{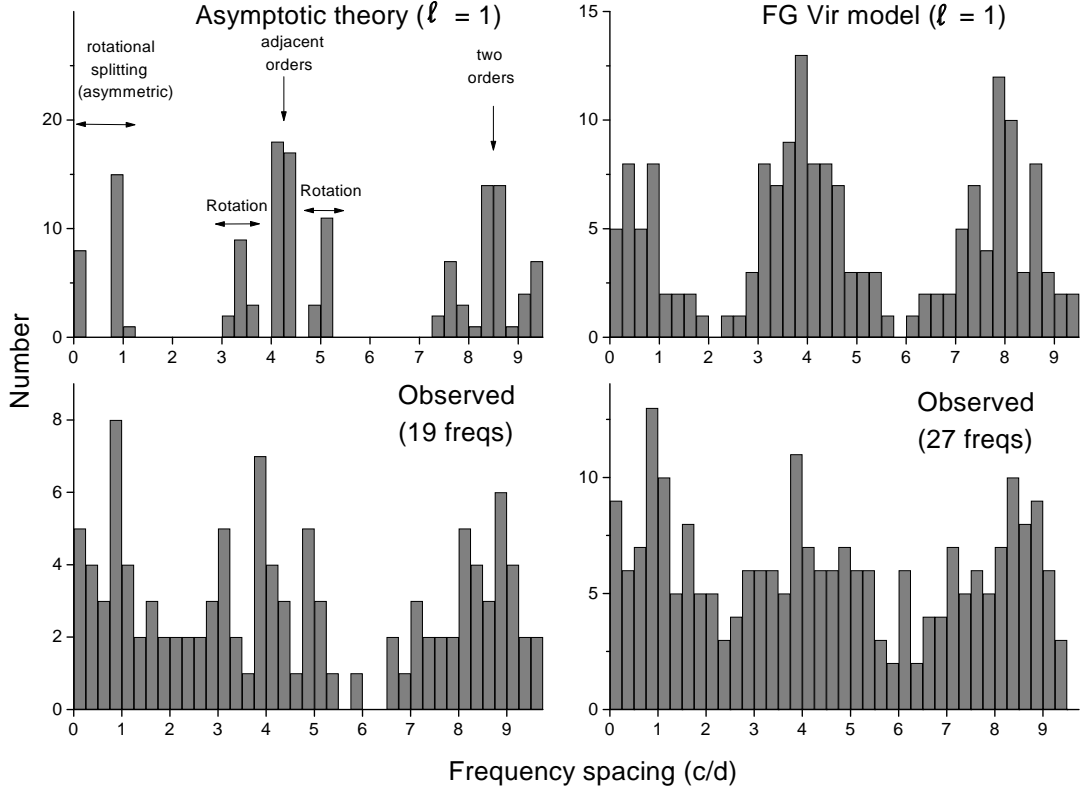}
\caption{
Histograms of frequency spacing between all specified pulsation modes.
{\em Top left:} The diagram demonstrates that for high orders the patterns
of frequency spacing clearly show adjacent radial orders ($\sim$ 4~c/d)
and the effects of rotational splitting, which is extremely asymmetric even
at $V_{\rm rot} = 45$~km/s.
{\em Top right:} The frequency spacing predicted from model~1
for $\ell = 1$ in the observed frequency range of 11 -- 35~c/d.
Note that the patterns from adjacent orders and rotational
splitting are still present. {\em Bottom panels:} Observed frequency
spacings in the observed range from 11 to 35~c/d. Although these are
a mixture of $\ell$ = 0, 1 and 2 modes, the effects of adjacent
radial orders and a small peak in the range of rotational splitting 
can be seen. 
To demonstrate that the results are not sensitive to which observed
frequencies are included, two different choices (see text) are shown}
\end{figure*}

\section{Regularities of frequency spacing}

The values of the observed frequencies and regularities in their
patterns can be an excellent intital tool for mode identifications,
if enough frequencies are excited and detected.
For high-order, low-degree p-mode pulsation,
the different radial orders show uniform frequency spacing,
with a mode of order $n$ and of degree $\ell$ being shifted from the
corresponding mode ($n$, $\ell + 1$) by half of the frequency difference
between the ($n$,$\ell$) and ($n+1$,$\ell$) modes (Vandakurov 1967).
In $\delta$~Scuti stars, the excited pulsation modes are of low order
($n$ up to 7), so that the asymptotic relations do not apply exactly.
Nevertheless, they also show some regularities.
Additionally g-modes invade the p-mode region and decrease
the spacing in a small frequency region of about two radial orders.
This effect, known as avoided crossing
(Osaki 1975, Aizenman, Smeyers \& Weigert 1977), complicates the theoretical
frequency spectra, but can provide information about the stellar interior
(Dziembowski \& Pamyatnykh 1991).
Moreover, stellar rotation splits multiplets
and this splitting is non-symmetric, if second-order effects of rotation
and effects of rotational mode coupling are taken into account
(Dziembowski \& Goode 1992, Soufi et al. 1998, Pamyatnykh et al. 1998).
Nevertheless, the spacing of adjacent radial orders as well as
the rotational splitting is still regular enough to be detectable, if
complete multiplets are excited and identified.
We will demonstrate this by using a pulsation model of a 1.85$\, M_\odot$
star with $T_{\rm eff} = 7515\, $K, $\log \, g = 3.99$, and 
$V_{\rm rot} = 45\, $km/s. This model will be referred to as model~1.
The parameters for the model were not chosen at random,
but can be regarded as an estimate for FG~Vir.

To investigate the period regularities, Winget et al. (1991) have successfully
applied the method of the Fourier transform of the period spacing to the star
PG~1159+035. This method requires coherence over a large frequency range.
Handler et al. (1997) also found frequency regularities from Fourier
transformations of the frequency spectrum of the unevolved $\delta$~Scuti
star XX~Pyx. Since strict equidistant frequency or period spacing is not
expected for FG~Vir, the method is not optimal for this $\delta$~Scuti star.
Instead, we use a method which does not require such a coherence: an
examination of a histogram of the observed frequency differences between all
detected frequencies. 
In such a diagram, regularities in the frequency spacing of adjacent radial
orders of modes with the same degree,
$\ell$, should show up as a peak. Furthermore, modes of different
degree are shifted in frequency relatively to each other, but would still
have similar patterns and, therefore, contribute to the peaks in the histogram.

The frequency spacing is examined in Fig.~1 with both the theoretically
predicted and observed spacings. Pulsation models show a typical frequency
spacing of $\Delta f \approx 4\, $c/d for adjacent radial orders of p-modes,
independent of the degree of the modes. 
The leftmost peaks in the top panels of Fig.~1 
are caused by rotationally split multiplets. A similar diagram for $\ell = 2$
(not plotted separately) does not show such strong peaks in the expected
region. The reason is that both the presence of g-modes in addition to the
p-modes and non-equidistant rotational splitting significantly disturb the
regularity in the distribution of quadrupole mode frequencies
(see Fig.~7 below.) As a result, the combined pattern
of frequency spacings for all $\ell = 0 - 2$ modes becomes much less clear.
Moreover, due to the fact that only low-order oscillations are present
in this frequency range, there is no additional peak at
$\Delta f \approx 2$ c/d as might be expected from the asymptotic
spacing between p-modes of adjacent degrees (see Fig.~7 for more details).

Next, we turn to the observed frequency spacing for the 24 certain and
8 probable frequency detections of FG~Vir (Table~1). The most
cautious approach would be to use the 24 certain frequencies
with a few exceptions: the 2$f_1$ term at 25.4~c/d (reflecting
the departure from a pure sinusoidal light curve shape of $f_1$), the
two combination frequencies (the pulsation models cannot yet predict
which combinations and resonances are excited), and the two low-frequency
modes for which the p-mode character can definitely be excluded 
from the assumption that $f_6$ is the radial fundamental mode.
To show that the agreement between the theoretically
predicted and observed frequency spacing is not based on the choice
of frequencies, the analysis was repeated by including the 8 additional
'probable' modes listed in Table 1.

To conclude, the theoretical and observed frequency spacings agree quite
well. In particular, for frequency differences in the 0 -- 5~c/d range,
two features near 3.9 and 0.8~c/d stand out, suggesting an
identification with the spacing of successive radial orders and
rotational splitting, respectively.

% *********************************************************************
\section{Pulsation mode identifications from photometric phase differences}

The relative phase difference between the temperature
and radius variations of a pulsating star leads to an observable
phase difference between the light curves at different wavelengths.
The sizes of these phases differences depend not only on the properties
of the star, but also on the type of pulsation mode. The observed
phase difference can then be used for mode typing. This was
already pointed out by Watson (1988).
Garrido et al. (1990) presented detailed calculations and predictions
for $\delta$ Scuti stars. They find that measurements through different
filters of the
Str\"omgren $uvby$ system provide discrimation between radial and low-order
nonradial pulsation, i.e. help determine the $\ell$ value \footnote
{It is necessary to note that rotational mode coupling
may enlarge the overlapping between
modes of different $\ell$ values in the amplitude-phase diagrams, as it was
discussed by Pamyatnykh et al. (1998):
for example, a quadrupole mode coupled by rotation with the
closest radial mode may be shifted in such a diagram
towards the region occupied by dipole modes.}.

\begin{table*}
\caption{Phase differences and mode identifications of FG Vir}
\begin{tabular}{lcccccc}
\hline
\noalign{\smallskip}
\multicolumn{3}{c}{Frequency}& \multicolumn{2}{c}{Phase differences
in degrees} & \multicolumn{2}{c}{ Pulsation degree, $\ell$}\\
& c/d & $\mu$Hz& \multicolumn{2}{c}{$\phi_v - \phi_y$} & Spectroscopy & Photometry\\
& & & 1995 & 1995/6 & Viskum et al. (1998) & Present\\
\noalign{\smallskip}
\hline
\noalign{\smallskip}
$f_1$ & 12.716 & 147.2 & --1.0 $\pm$ 0.2 & --1.3 $\pm$ 0.2 & 1 & 1\\
$f_2$ & 24.228 & 280.4 & --3.0 $\pm$ 1.1 & --3.8 $\pm$ 1.1 & 1 & 1, 2\\
$f_3$ & 23.403 & 270.9 & --0.6 $\pm$ 1.2 & --1.1 $\pm$ 1.2 & 0 & 0, 1\\
$f_4$ & 21.052 & 243.7 & --5.7 $\pm$ 1.4 & --7.1 $\pm$ 1.5 & 2 & 2\\
$f_5$ & 19.868 & 230.0 & --4.9 $\pm$ 1.4 & --5.5 $\pm$ 1.5 & 2 & 2\\
$f_6$ & 12.154 & 140.7 & +6.8  $\pm$ 1.4 & +5.5  $\pm$ 1.4 & 0 & 0\\
$f_7$ & 9.656 & 111.8 &  --2.2 $\pm$ 1.4 & --2.7 $\pm$ 1.4 & 2 & 1, 2\\
$f_8$ & 9.199 & 106.5 &  --4.4 $\pm$ 1.6 & --7.6 $\pm$ 1.7 & 2 & 2\\
\noalign{\smallskip}
\multicolumn{3}{l}{Number of hours $y/v$}& 412/292 & 494/374\\
\noalign{\smallskip}
\hline
\end{tabular}\newline
\end{table*}

\begin{figure}
\centering
\includegraphics*[bb=24 30 279 510,width=88mm]{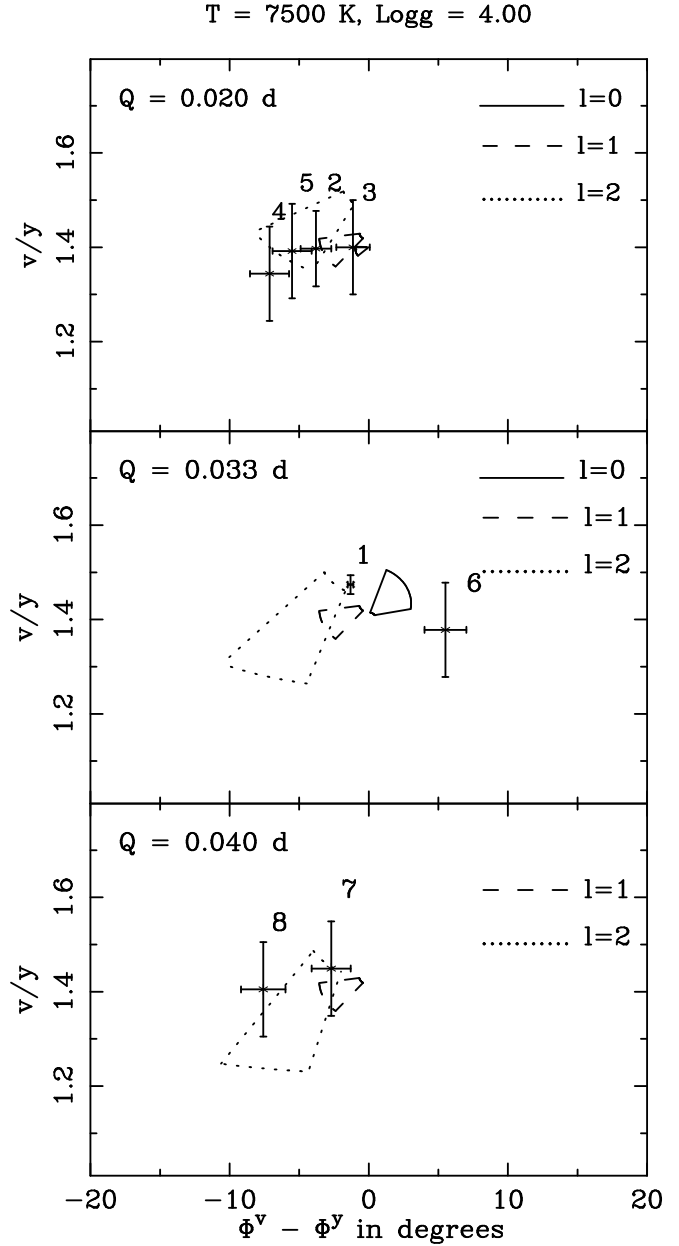}
\caption{Diagnostic diagram to determine $\ell$ values of FG Vir from
Str\"omgren $v$ and $y$ colors. The axes represent amplitude ratios and phase
differences. Measurements are shown by crosses with error bars, while
the four-sided loops represent the models (see text).
The three panels represent the pulsation modes with different values
of the pulsation constant, $Q$}
\end{figure}

We have chosen the $v$ and $y$ filters to provide a relatively large baseline
in wavelength. The $u$ filter was not used by us because of the very large
potential for systematic observational errors. These details of
the measurements
can be found in Breger et al. (1998). The phase differences were determined
in the following manner: The values of the 24 known and well-determined
frequencies were optimized by making a common
solution of the available $y$ data from 1992 - 1996, while allowing
for the amplitude
variability of $f_3$. As discussed in Breger et al., all CCD measurements
were given a
weight of 0.19. With these optimized frequencies, for the year 1995
the best amplitudes
and phases were calculated from the available 412 hours
of $y$ and 292 hours of $v$ data.
Separate trial solutions indicate that the resulting phase differences
are relatively insensitive to the weights adopted. For the year 1996,
an additional 82 hours
of $uvby$ photometry are available (Viskum et al. 1998). The data were
combined with the larger data set from 1995 while allowing for
variable amplitudes of $f_3$ .We note
that the calculated uncertainties of the phase differences
are not reduced by including the additional data: the reason
is that the 1995 data have smaller deviations, e. g. 4 vs. 6 mmag
in $v$.

\begin{figure}
\centering
\includegraphics*[bb= 82 37 725 511, width = 87mm]{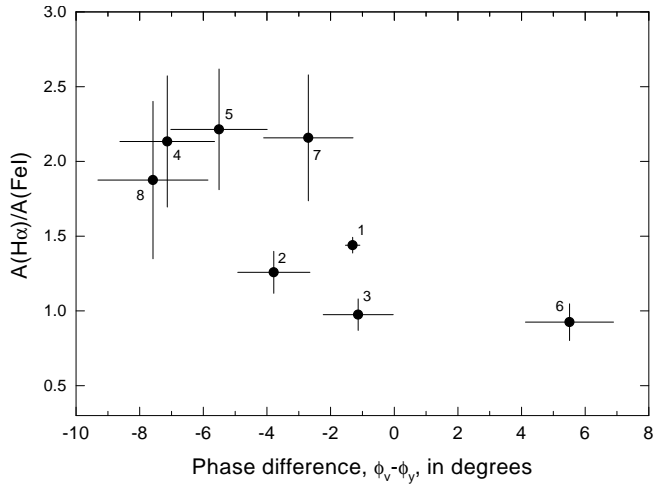}
\caption{Comparison of the equivalent-width and photometric methods
to determine
$\ell$ values. Radial pulsation ($\ell$=0) can be found in the lower right,
$\ell$=1 in the middle, while $\ell$=2 is found near the top left.
The diagram shows that the two methods are in agreement, but also demonstrates
that the some of the agreement may be accidental once the error bars
are taken into consideration}
%\end{figure*}
\end{figure}

The resulting phase differences are shown in Table 2.
The phase errors (in degrees)
were estimated from the formula $\sigma(\phi) = 180/\pi \cdot \sigma (m)/a$,
where $a$ is the amplitude and $\sigma (m)$
is the uncertainty of each data point (average deviation
per point from the fit).
 
We can now compare the observed phase differences with theoretical
modelling in order to determine the $\ell$ values. The ATLAS9 models of
Kurucz (1993) were used to construct a model atmosphere for FG Vir.
Garrido et al. (1990) presented calculations using values of $\Phi^T$
ranging from 90$\deg$ to 135$\deg$. For FG Vir, Viskum (1997) determined a
smaller range, viz. $\Phi^T = 126\deg \pm 20\deg$. This allowed us to
refine the calculations, although the results are  very similar.
Another required constant, the deviation from adiabaticity, R,
has been changed slightly from the value used by Garrido et al. (1990).
Values of 0.20 (instead of 0.25) to 1.00 were used. This change was
indicated by measurements of high-amplitude $\delta$ Scuti stars.
The theoretical predictions are shown in Fig.~2 together with the
observations. The importance of considering the dependence on the
pulsation constant, $Q$, can be seen for $Q$ = 0.02, where one can
even find negative values of the phase difference for radial modes,
although the separation between radial and nonradial modes is always maintained.

Our best mode identifications based on Str\"omgren photometry are shown in Table 2.
We obtained five unambiguous $\ell$ values, while for three further modes we cannot
distinguish between two adjacent $\ell$ values. The frequency $f_6$, shown in the
middle panel, is situated to the right of the $\ell$ = 0 by 1.9 $\sigma$.
We note that the deviation is caused by only one subset of data (the CCD
measurements from Siding Spring Observatory, see Stankov et al. 1998),
without which a phase shift of +3.7 $\deg$ is found. Irrespective of which
of the two values for $f_6$ is accepted, an identification with radial pulsation is
consistent within the statistical uncertainties.

We can now compare the results from the photometric method
with those derived from a promising new technique of examining
the equivalent width variations of selected lines.
Bedding et al. (1996) have shown that
for low degree pulsation, the $\ell$-values of pulsation modes
can be inferred from simultaneous observations of several selected
absorption lines combined with simultaneous photometric observations.
Viskum et al. (1998) have applied this method to the star FG~Vir.
In particular, the equivalent-width changes of the H$\alpha$ and
H$\beta$ lines turned out to be good discriminators.
In their paper, $\ell$ identifications have been presented
for the eight dominant modes.

We note that on the observational side the photometric and
spectroscopic methods are independent. However, both methods
rely on similar model-atmosphere calculations, so that
they cannot be considered to be completely independent of each other.

The agreement between the photometric and spectroscopic mode determinations
is remarkable. It appears prudent to examine the comparison of the
results of the two methods in more detail, especially with
consideration of the (unavoidable) observational uncertainties.
In order to compare independent parameters with each other,
we pick the amplitude ratio of
A(H$\alpha$)/A(FeI) given by Viskum et al. (1998).
The comparison is shown in Fig.~3, where the numbers next to the
points refer to the frequency numbering in Table 1. The figure shows
that some of excellent agreement may be accidental once the
observational uncertainties are considered. Nevertheless, the
viability of both methods to determine $\ell$ values has been
demonstrated and for at least six modes the $\ell$ values have
been observationally determined. These determinations now need
to be used as input for pulsation models.

\section{Pulsation models for FG~Vir}

Since the initial discovery of multiperiodicity of FG Vir, several studies
attempted to fit the observed and theoretical frequency spectra of
the star, viz. Breger et al. (1995), Guzik, Templeton \& Bradley (1998),
and Viskum et al. (1998). We will now calculate new models utilizing the
newly discovered pulsation frequencies and mode identifications.

\subsection{Method of computation}

To compute models of FG~Vir we used a standard stellar evolution code
which was developed in its main parts by
B.~Paczy\'nski, M.~Koz{\l}owski and R.~Sienkiewicz (private communication).
The same code was used in our recent studies
of period changes in $\delta$~Scuti stars (Breger \& Pamyatnykh 1998)
and in a seismological study of XX~Pyx (Pamyatnykh et al. 1998).
These two papers include detailed descriptions of the model computations,
so that the present decription can be brief.
For the opacities, we used the latest version of the OPAL or the OP tables
(Iglesias \& Rogers 1996 and Seaton 1996, respectively)
supplemented with the low--temperature data of Alexander \& Ferguson (1994).
In all computations the OPAL equation of state was used (Rogers et al. 1996).

The computations were performed starting with
chemically uniform models on the ZAMS,
assuming typical Population I values of hydrogen
abundance, $X$, and heavy element abundance, $Z$. The initial heavy element
mixture of Grevesse \& Noels (1993) was adopted. 

In some models, a possibility of overshooting from the convective core 
was taken into account.
The overshooting distance, $d_{\rm over}$, was chosen
to be $0.2 \, H_{\rm p}$,
where $H_{\rm p}$ is the local pressure scale height at the edge
of the convective core. 
Examples of evolutionary tracks for $\delta$~Scuti
models computed with and without overshooting are given
in Breger \& Pamyatnykh (1998).

In the stellar envelope the standard mixing-length theory of convection
with the mixing-length parameter $\alpha$ = 1.0 or 2.0 was used.
As we will see below, the choice of the mixing-length
parameter $\alpha $ has only a small effect on our models, because they are too
hot to have an effective energy transfer by convection in the stellar
envelope. 

In all computations we assumed uniform (solid-body) stellar
rotation and conservation of global angular momentum during evolution
from the ZAMS. These assumptions were chosen due to their simplicity.
The influence of rotation on the evolutionary tracks of $\delta$~Scuti
models was demonstrated by Breger \& Pamyatnykh (1998).
We studied models of FG~Vir with equatorial rotational velocities from,
approximately, 30 to 90 km/s (on the ZAMS, the values are 5-10 km/s higher).
This range is consistent with the values of $v \sin i = 21 \pm 1\, $km/s
and $i = 31\deg \pm 5\deg$ found by Mantegazza et al. (1994) and an
equatorial velocity of $33 \pm 2\, $km/s obtained by Viskum et al. (1998).
At such low rotational velocities, the evolutionary tracks are located
very close to those for non-rotating stellar models. 
The main effect of rotation to be considered 
is the splitting of multiplets in the oscillation frequency spectra.
This splitting is non-symmetric even for slowly rotating stars, if
second-order effects are included. 

The linear nonadiabatic analysis of low-degree oscillations ($\ell$\,$\leq$\,4)
was performed using the code developed by Dziembowski (1977).
In the modern version of the code, effects of slow stellar rotation
on oscillation frequencies are taken into account up to second order
in the rotational velocity (Dziembowski \& Goode 1992, Soufi et~al. 1998).

% *********************************************************************
\subsection{Model constraints using oscillation data}

The models for FG~Vir were constructed with the observed mode
$f_6$ (12.154~c/d)
being identified with the radial fundamental mode (={\bf F}) (see Section~3).
Note that this determines the mean density of
all possible models of FG~Vir: with the pulsation constant
of about 0.032 -- 0.034~days,
which is typical for $\delta$~Scuti variables, we obtain
$\bar\rho/\bar\rho_{\odot} \approx$ 0.15--0.17.
A considerably more accurate value of the density will be obtained
later in this section.

We started with the construction of evolutionary tracks of 1.75 -- 1.95
$M_{\odot}$ models for initial abundances $X=0.70$ and $Z=0.02$
and using OPAL opacities. No overshooting from the convective core was allowed.
The initial equatorial rotational velocity on the ZAMS was chosen
to be 50 km/s.
With our assumption of conservation of global angular momentum,
the equatorial rotational velocity is decreasing during the
MS--evolution from 50 km/s at the ZAMS to about 40--41 km/s at the TAMS
(Terminal--Age--Main--Sequence).
The evolutionary tracks are shown in Fig.~4 together with the range in 
effective temperature and gravity of FG~Vir (see Introduction)
derived from photometric calibrations
($T_{\rm eff} = 7500 \pm 100$~K and $\log \, g= 4.00 \pm 0.10$).
This range requires MS models and constrains the mass of the models to 
1.75--1.95~$M_{\odot}$. The position of the models, whose {\bf F}-mode
has a value of 12.154~c/d is also shown. This further constrains
the mass to 1.815--1.875~$M_{\odot}$. We stress that this strong seismological
mass constraint depends on an accurate effective temperature determination.

The identification of $f_6 \equiv $ {\bf F} 
agrees well with the gravity estimate for FG~Vir. This provides no
additional constraints on the stellar mass since the
lines of constant frequencies are approximately parallel
to those of constant gravity, as shown in the lower panel of Fig.~4.

For a given family of stellar models, the radial fundamental pulsation
constant, $Q$, is constant with a quite high accuracy due to the homologous
structure of models of different masses.
For $M$ = 1.80 -- 1.90 $M_{\odot}$ in the range
$\log \, T_{\rm eff}$ = 3.869 -- 3.881 ($T_{\rm eff}$ = 7400 -- 7600 K),
the pulsation constant $Q$ = 0.0326 with relative accuracy of about 0.2 \%.
The accuracy will be still higher by approximately one order of magnitude
if we consider only models based on $f_6 \equiv $ {\bf F}. 
For these models we determine a mean
density of $\bar{\rho}/\bar{\rho_{\odot}} = 0.1570 \pm 0.0001$. 
However, such an extremely high accuracy is based on
a fixed choice of input physics: stellar opacity, 
initial chemical composition, rotational velocity and parameters of convection.
We will see in the next subsection that changing these parameters
results in significantly larger spread of mean density for FG Vir models.

\begin{figure}
\centering
\includegraphics*[width=88mm]{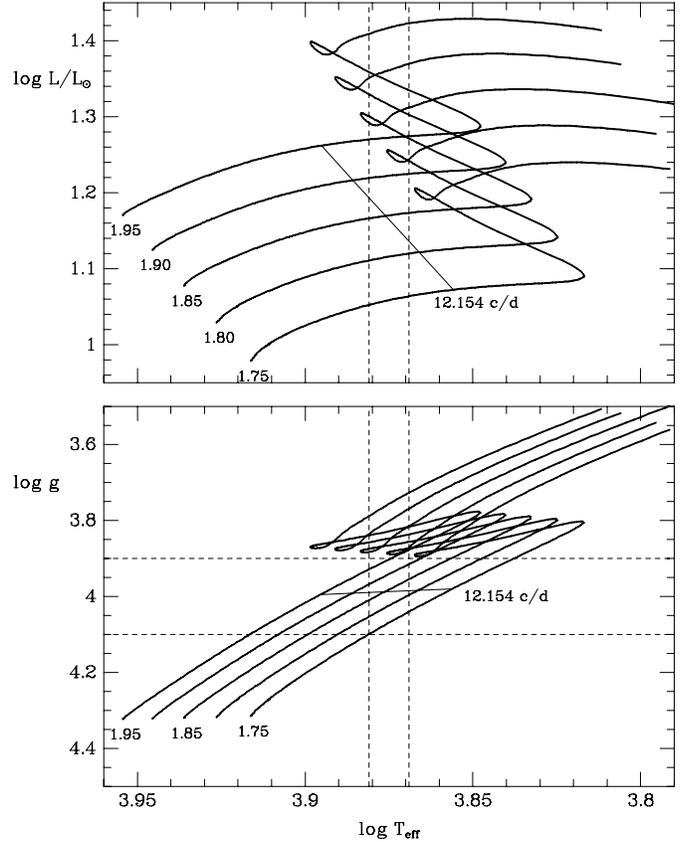}
\caption{Evolutionary tracks of 1.75--1.95 $M_{\odot}$ standard models. 
The equatorial rotational velocity
on the ZAMS was chosen to be 50~km/s. On the TAMS
(at the turn-off points to the left),
equatorial velocities are of about 40--41~km/s.
Dashed lines show effective temperature and $\log g$ ranges
of FG~Vir from photometric calibrations.
The thin solid line connects models whose radial fundamental mode 
frequency is 12.154~c/d.
The rotational velocity of the models along this line is about 45~km/s}
\end{figure}

The strict constraint on mass is demonstrated in Fig.~5, where the
changes of radial and dipole frequencies during MS evolution are shown.
This agreement is an independent qualitative argument in favour
of the proposed models for FG~Vir.
Another important result is the good agreement of the predicted
frequency range of unstable modes with the observed frequency range of
9--34~c/d.
An additional test shows that should we identify $f_6$
with the first radial overtone instead of the {\bf F}-mode, we cannot
achieve agreement between the theoretical and observed frequency ranges:
in the corresponding models of $\approx 2.0\, M_{\odot}$ the instability
occurs in the frequency range of 8--30~c/d. The tendency in models of higher
mass to shift the instability range to lower frequencies
can be also seen in Fig.~5. There is an even stronger argument
against these higher-mass models: their luminosities are too high
($\log (L/L_{\odot}) \sim 1.3$)
to be consistent with both the photometric calibrations and the
Hipparcos parallax.

\begin{figure*}
\centering
\includegraphics*[width=150mm]{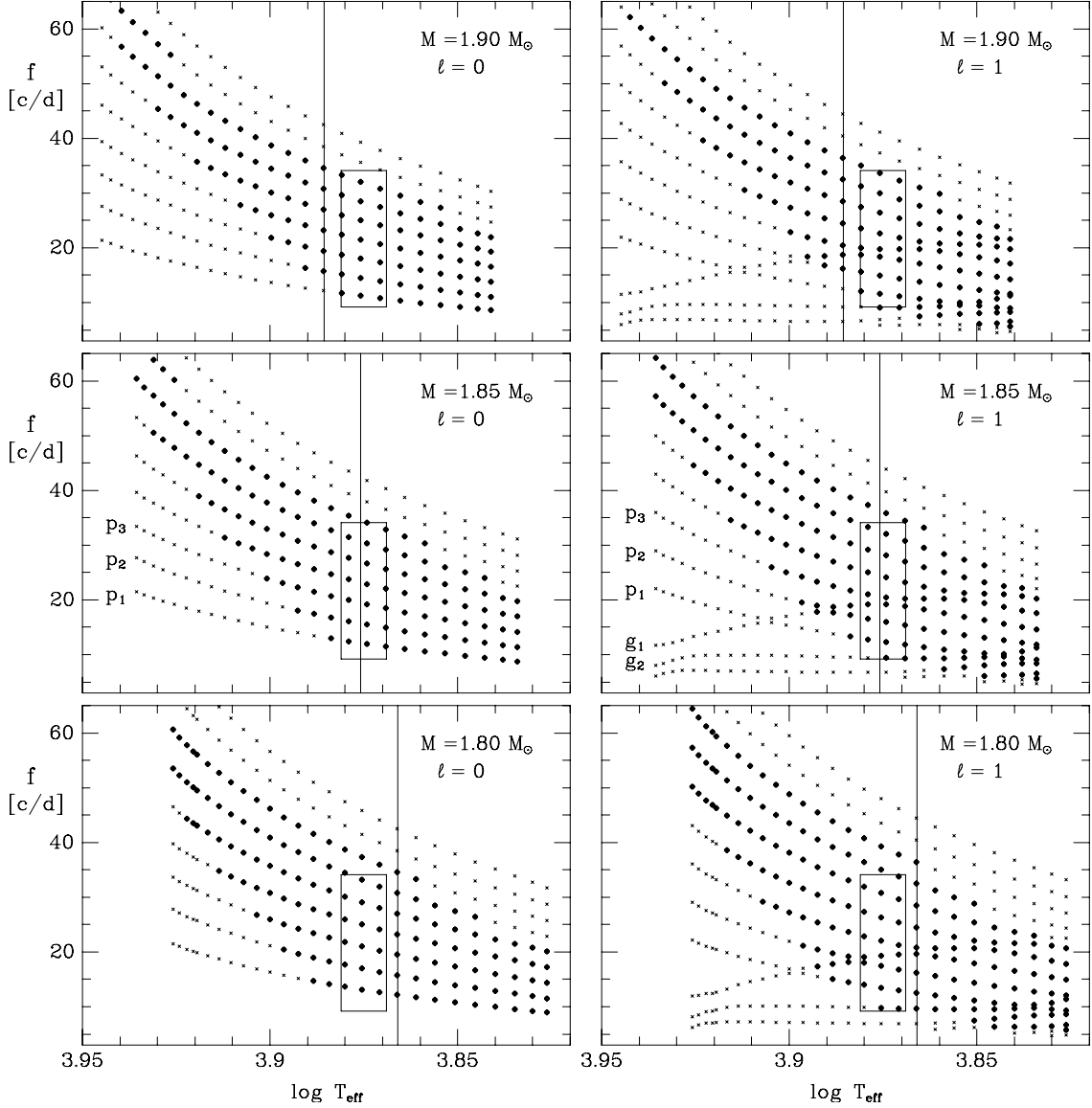}
\caption{Main-sequence evolution of low-order frequency spectra of radial
and dipole oscillations of stellar models with masses 1.80, 1.85
and 1.90~$M_{\odot}$.
In each panel, lefmost and rightmost points correspond to the ZAMS and
to the TAMS models, respectively.
Large filled circles denote unstable modes.
For simplicity, for $\ell = 1$ only axisymmetric ($m = 0$) components
of the dipole multiplets are shown.
Rectangular boxes mark the observational frequency  and
effective temperature range of FG Vir. 
The vertical line in each panel denotes
a model whose radial fundamental frequency ({\bf F}-mode) fits
the observational frequency $f_6$ = 12.154 c/d.
Only models with masses 1.815--1.875~$M_{\odot}$
fit the allowed temperature range}
%  \label{hrd}
\end{figure*}
%\end{figure}

A number of gravity modes must be excited in FG~Vir, 
if the assumption of $f_6 \equiv $ {\bf F} is true, because the two lowest
frequencies are more than 25\% lower than the {\bf F}-mode. 
Moreover, during the MS-evolution the frequencies of low-order g-modes
increase and approach consecutively 
the frequencies of low-order
p-modes resulting in mode interactions and avoided crossings
(see Unno et al. 1989 and references therein). The frequency spectrum
is much more complicated than in the case of pure p- or g-modes, as 
shown in Fig.~5 for dipole modes.
The avoided crossing phenomenon takes place approximately in the middle of
the observed frequency interval. Therefore, most of the excited modes at
these and at lower frequencies are of mixed character: they behave like
p-modes in the envelope and like g-modes in the interior. In the 
1.85~$M_{\odot}$ model with
$\log \, T_{\rm eff} = 3.876$, modes $g_1$, $p_2$ and $p_3$ are of
mixed character. The frequencies of modes at avoided crossing are sensitive
to the structure of the deep stellar interior. Consequently, 
the detection of these modes is important for testing convective
overshooting theories (Dziewbowski \& Pamyatnykh 1991).

Avoided crossings for quadrupole modes in the models of FG~Vir
occur close to the upper border of the observed frequency interval and also
close to the {\bf F}-mode. 
This means that most of $\ell =2$
p-modes in the interval already interacted with gravity modes and are of
mixed character.

% *********************************************************************
\subsection{Effects of different input parameters on the FG~Vir models}

The 1.85 $M_{\odot}$ model for FG~Vir, which was discussed in the
previous subsection, will be referred to as the standard or reference model
with the input parameters: $X=0.70$, $Z=0.02$, $d_{\rm over}=0$,
$\alpha = 1.0 $, $V_{\rm rot,ZAMS}=50 $~km/s and
OPAL opacities.
To examine the effects of varying input parameters on the predicted
frequency spectrum, all these and the stellar mass were varied, 
under the condition that ${\bf F} \equiv f_6$. 

% *********************************** Table 3.
\begin{table*}
\begin{center}
\caption{Parameters of FG~Vir models with ${\textbf F} \equiv f_6$.
The symbols have their usual meaning (see text). For the opacity,
$\kappa$, the OPAL, OP or arficially
modified OPAL data were used.
$p_1 / p_4$ is the ratio of frequency of radial fundamental mode, $f(p_1)$,
to that of third overtone, $f(p_4)$}
% \label{models}
\begin{tabular}{|cccccccccccccc|}
\hline
Model & $M/M_{\odot}$ & $X$ & $Z$ & $\log T_{\rm eff}$ & $\log L$
& $R/R_{\odot}$ & $\log \, g$ & $V_{\rm rot}$ & $\alpha$ & $d_{\rm over}$
& $\kappa$ & $\bar\rho/\bar\rho_{\odot}$ & $p_1 / p_4$\\
\hline
 1 & 1.85 & 0.70 & 0.02 & 3.8760 & 1.1690 & 2.274 & 3.988 & 45  & 1.0 & 0.0 &      OPAL & %%@
0.1571 & 0.5236\\
 2 & 1.82 & 0.70 & 0.02 & 3.8701 & 1.1406 & 2.261 & 3.985 & 45  & 1.0 & 0.0 &      OPAL & %%@
0.1571 & 0.5233\\
 3 & 1.85 & 0.70 & 0.02 & 3.8761 & 1.1696 & 2.274 & 3.989 & 31  & 1.0 & 0.0 &      OPAL & %%@
0.1560 & 0.5231\\
 4 & 1.85 & 0.70 & 0.02 & 3.8756 & 1.1676 & 2.274 & 3.983 & 67  & 1.0 & 0.0 &      OPAL & %%@
0.1570 & 0.5248\\
 5 & 1.85 & 0.70 & 0.02 & 3.8753 & 1.1656 & 2.272 & 3.976 & 90  & 1.0 & 0.0 &      OPAL & %%@
0.1563 & 0.5266\\
\hline
 6 & 1.85 & 0.70 & 0.02 & 3.8760 & 1.1691 & 2.274 & 3.988 & 45  & 2.0 & 0.0 &      OPAL & %%@
0.1570 & 0.5236\\
\hline
 7 & 1.85 & 0.70 & 0.02 & 3.8796 & 1.1837 & 2.275 & 3.987 & 45  & 1.0 & 0.2 &      OPAL & %%@
0.1558 & 0.5231\\
\hline
 8 & 1.85 & 0.65 & 0.03 & 3.8734 & 1.1562 & 2.267 & 3.990 & 45  & 1.0 & 0.0 &      OPAL & %%@
0.1584 & 0.5227\\
\hline
 9 & 2.00 & 0.70 & 0.03 & 3.8748 & 1.1844 & 2.327 & 4.001 & 46  & 1.0 & 0.0 &      OPAL & %%@
0.1574 & 0.5231\\
\hline
10 & 1.72 & 0.70 & 0.02 & 3.8754 & 1.1507 & 2.233 & 3.972 & 45  & 1.0 & 0.0 &      OP   & %%@
0.1542 & 0.5262\\
\hline
11 & 1.95 & 0.70 & 0.02 & 3.8712 & 1.1600 & 2.301 & 4.000 & 45  & 1.0 & 0.0 &  mod.OPAL & %%@
0.1588 & 0.5204\\
12 & 1.95 & 0.70 & 0.02 & 3.8746 & 1.1738 & 2.301 & 4.002 & 32  & 1.0 & 0.2 &  mod.OPAL & %%@
0.1597 & 0.5195\\
\hline
\end{tabular}
\end{center}
\end{table*}

The changes introduced by using different opacities or non-standard chemical
composition were mainly compensated by changes in mass, in order 
to fulfill the only identification we use. 

The main characteristics of twelve models of that series are given in Table~3.
Model 2 differs from model 1 (our reference model) in mass;
models 3, 4, 5 - in rotational velocity; model 6 versus 1
will demonstrate effect of changing the mixing-length parameter $\alpha$;
model 7 versus 1 will show effect of the overshooting;
models 8 and 9 have non-standard chemical composition;
finally, models 10, 11 and 12 differ from model 1 in opacity (additionally,
overshooting is taken into account in the model 12).

Note the significantly larger spread in stellar mass between different
models (1.72--2.00 $M_{\odot}$) than for the mass interval
of 1.815--1.875 $M_{\odot}$ in the case of the standard choice of input
parameters as was discussed in the previous subsection.
The same is true for the mean density range:
in Table~3 it varies between $\bar\rho/\bar\rho_{\odot}$ = 0.1542 and 0.1597
(or between 0.1558 and 0.1584 when using only OPAL opacities). This spread
is at least one order of magnitude larger than for the standard input data.
Nevertheless, this seismic estimate of the mean density, which is based both on
the well determined effective temperature and the one identification we are
using, provides a strong constraint on possible FG Vir models \footnote
{For the multiperiodic $\delta$~Scuti-type star XX~Pyxidis, for example,
there is no observational information about mode identification.
Therefore it was necessary to consider a large number of models with very
different mean densities (Pamyatnykh et al. 1998).
}.

We note that besides quite different stellar masses of 1.7 -- 2.0 $M_{\odot}$
(see Table~3) the evolutionary tracks for all 12 models in their MS-part 
lie well inside the region of 1.80 -- 1.90 $M_{\odot}$ of the standard set.
Including a luminosity estimation $\log (L/L_{\odot}) \approx  1.1 - 1.2$ 
from trigonometric parallax determined by Hipparcos, all MS model
tracks pass the error box, as well as the error box in the 
$\log  g$-$\log T_{\rm eff}$-diagram.
On the contrary, none of the post-MS
models fits such a combination of parameters. 

% *********************************************************************
\subsection{The problem of the radial frequency ratio}

Viskum et al. (1998) identified $f_3$ as a radial mode. We note here
that the phase-difference method presented earlier in this paper allows
both $\ell$ = 0 and 1 for $f_3$, i. e. radial as well as nonradial pulsation. We
will now examine the radial hypothesis.
In the last column of Table~3 the ratio of frequencies of the radial
fundamental mode, $p_1$, and  of the third overtone, $p_4$, is given
($f(p_1)/f(p_4)$). 
For models 1--10 these values are close to, but not equal to the observed
ratio of $f_6 / f_3 = 0.5193$, independent of which parameter was changed. 
This can also be seen in Fig.~6, where the ratio $f(p_1)/f(p_4)$, is plotted
against $f(p_1)$ for a wide range of parameters of $\delta$~Scuti star models. 
There are well-defined
monotonic variations of this ratio with changing mass, 
effective temperature or chemical composition, but the observed ratio
disagrees with all these results. 

\begin{figure*}
\centering
\includegraphics*[width=130mm]{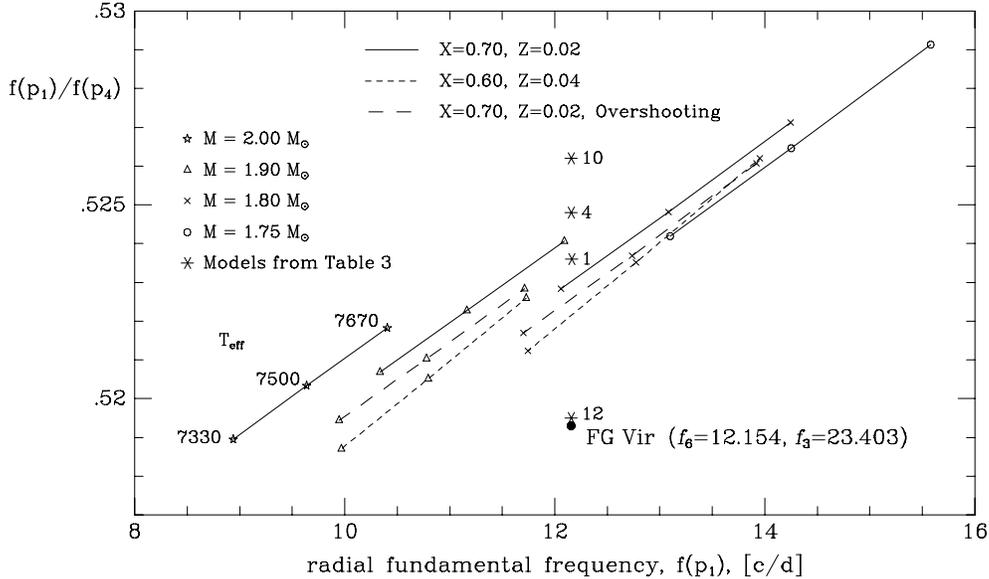}
\caption{Frequency ratio of the radial fundamental mode to the third overtone
for a wide range of parameters of $\delta$~Scuti star models and 
of some FG~Vir models from Table~3 (asterisks). 
The large filled circle corresponds to the observed frequency
ratio, $f_6 / f_3$, of 0.5193. 
Only the models with artificially modified
opacities (such as model 12 of Table~3) can fit the observed ratio}
\end{figure*}

The only exception is the model 12 with artificial opacities,
which was constructed in the following way. For FG Vir models, the frequency
ratio is most sensitive to the choice of opacities as it can be seen from
comparison of models 10 and 1 (OP versus OPAL opacities). Physically,
OP data differ from those of OPAL by underestimation of collective effects
in stellar plasma, therefore OP opacities are systematically lower than
OPAL in deep stellar interiors. For FG~Vir models,
this difference in opacity is about 20\% at temperatures above
$10^6$~K. In the envelope, at lower temperatures, OP opacity varies
slightly more monotonously along radius than does OPAL opacity: some dips
are slightly shallower and some bumps are more flat. The differences
do not exceed 8\%: for example, at a temperature of $14\, 000$~K the OP
opacity is 4\% smaller and at a temperature of $300\, 000$~K it is 7.5\%
higher than the OPAL opacity. 

Using the fact that the difference in frequency ratios between model 1 (OPAL)
and model 10 (OP) is comparable with the difference between model 1 and the
observations (but these differences are of opposite sign, see Fig.~7),
we performed a very simple numerical experiment: we artificially scaled
OPAL opacities with a factor, which is the ratio of OPAL to OP data.
More clearly, we used
$\kappa_{\rm modified} = \kappa_{\rm OPAL}  \cdot
[\kappa_{\rm OPAL} /\kappa_{\rm OP}] $.
Models 11 and 12 were constructed using $\kappa_{\rm modified}$.
For model 12 we additionally set $d_{\rm over} $ to $0.2\, H_p$
and lowered the rotational velocity.
This model fits the observed frequency ratio very nicely as demonstrated in
Fig.~7. However, this agreement should not be construed as an
indicator for a new revision of atomic physics data on opacity, since
the mode identification from Viskum et al. (1998) may not be unique due to
the size of the error bars. Moreover, we cannot exclude additional effects
like nonlinear mode interaction or rotational mode coupling, which may
influence the frequency spectrum. In the last section the problem of
rotational mode coupling will be briefly discussed. The observed variability
of the amplitude of mode $f_3$ is another argument in favour of
possible nonlinear mode interaction.

Viskum et al. (1998) were able to interpret the observed frequency ratio
$f_6$/$f_3 = 0.5193$ as the radial frequency ratio $f(p_1)/f(p_4)$. They did
not construct full evolutionary models but scaled a model 
of 2.2$ M_{\odot}$, which was selected to match the observed frequency ratio
with the radial frequency ratio $f(p_1)/f(p_4)$. Using the homology argument,
they estimated the mean density of the true FG~Vir model by
multiplying the mean density
of the 2.2$ M_{\odot}$ model by the square of the ratio 
$f_{\rm obs}$/$f_{\rm model}$. In such a way an agreement between observed
frequencies $f_6$, $f_3$ and a pair of radial modes of the scaled model
was achieved by definition. The estimated gravity, luminosity and distance
of the scaled model were found to be in good agreement with the photometric
and the spectroscopic data and with the Hipparcos 
parallax. The authors noted that the high precision of their
asteroseismic density estimate
($\bar{\rho}/\bar{\rho_{\odot}} = 0.1645 \pm 0.0005$) is based on a fixed
(solar) metallicity for FG~Vir. Indeed, with our standard choice of chemical
abundances and opacity and assuming the fit ${\textbf F} \equiv f_6$ 
we estimated the mean density of the FG~Vir model with even five times higher 
accuracy (see section 4.2), but possible variations of the global parameters
result in at least one order of magnitude worse precision of this estimate.
% *********************************************************************
\subsection{Theoretical frequency spectra versus observations}

Frequency spectra of radial, dipole and quadrupole modes for all 12 models
from Table~3 are shown in Fig.~7. The effects of different choices of input
parameters can be estimated by comparison of the results for different
models. We discuss here both general properties and some peculiarities of these
frequency spectra.

\begin{figure*}
\centering
\includegraphics*[width=180mm]{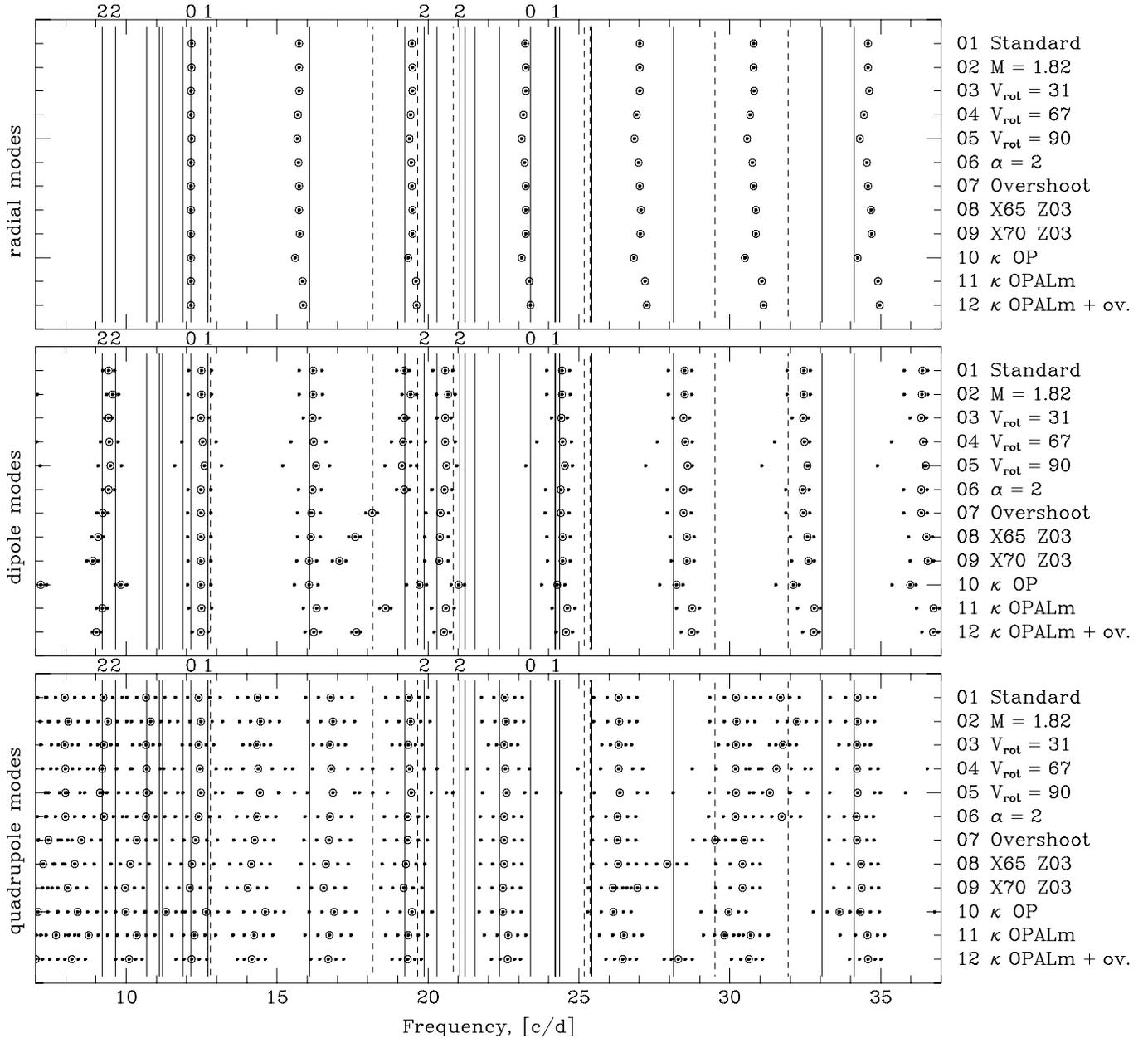}
\caption{Frequency spectra of radial, dipole and quadrupole
oscillations of various FG~Vir models.
Axisymmetric modes ($m=0$) are marked by enlarged circles.
Model numbers (see Table~3 for parameters) together with some model indicators
are given to the right of the panels.
Vertical solid and dashed lines correspond to observed frequencies --
statistically significant and probable, respectively.
Numbers above some observed frequencies give  identifications for the degree
of the modes ($\ell$) based on multicolor photometry data and on the results
by Viskum et al. (1998)}
\end{figure*}

For nonradial oscillations,
evolutionary overlapping of frequency intervals of g- and p-modes
(see Fig.~5) results in avoided crossings, which disturb the approximately
equidistant frequency spacing between acoustic multiplets. 
Gravity and mixed modes are very sensitive to the interior structure as can
be seen for models of different chemical composition, different opacities
and for models with and without overshooting. On the contrary, the change
of $\alpha$ (model 6 versus model 1) has a negligible influence on the
frequency spectrum due to ineffective convection in the relatively hot
envelope of FG~Vir. 

Rotation splits nonradial multiplets and strongly complicates the frequency
spectra. Except for models with slowest rotation, we observe a forest
of quadrupole modes in the low-frequency part of the interval,
with overlapping components of the different multiplets.
The common property of the spectra is a large asymmetry of the rotational
splitting, which is caused by the second-order effects of rotation 
(Dziembowski \& Goode 1992). 
The asymmetry is higher the higher the order of the p-modes is. 

It is not trivial to select a model reproducing the observed frequencies
exactly. Simple attempts to minimize frequency differences (O-C)
by a combined variation of input parameters of the reference model 
fail due to strong and non-linear sensitivity of gravity modes 
to interior structure. At the same time, this strong sensitivity may
help to fit some chosen frequencies without changing the rest of the spectrum,
(cf. models 1 and 7, for example). It is obvious from Fig.~7 that 
generally it is much easier to fit a low-frequency mode than
a high-frequency mode, because the spectrum is more dense at lower
frequencies. 
For some of the observed modes there is no satisfactory $\ell=0-2$ solution:
see, for example, the group $f_4$, $f_{14}$, $f_{18}$ around 21~c/d 
or $f_{17}$ at 33~c/d.
Besides geometric cancellation, there are no objections to identify
frequencies in the gaps with modes of degree $\ell$=3 and $\ell$=4. 
Note that even for $\ell=0-2$ the number of unstable modes is a few
times larger than the observed one: in the observational frequency interval
there are 6--7 radial modes, 24 dipole modes and 50--55 quadrupole modes.
Therefore, a presently unknown mode selection mechanism must exist.

Note that most of the models presented in Fig.~7 show a good fit
of the dominant observed mode $f_1$ (12.716 c/d) with $\ell = 1$ mode of
$m=-1$ or 0. This is in  agreement with the mode identification from photometric
phase differences. On the contrary, it is quite difficult to achieve a similar
fit with a dipole mode for the observed mode $f_3$ (23.403 c/d).
In Table~4 we present some results to quantify the fitting of
21~modes (corresponding to the observed frequencies $f_1$ through $f_{22}$,
omitting $f_{16}$) for models 1 (reference model), 3 (low rotation),
8 ($X=0.65$, $Z=0.03$) and 12 (artificially modified opacity + overshooting).
In the cases of close observed frequencies (for example, 
$f_1$, $f_{11}$, $f_{21}$ around 24~c/d, 
or $f_4$, $f_{14}$, $f_{18}$ around 21~c/d)
we give a few possible identifications for each frequency.
Moreover, if there is no $\ell=0-2$ mode for a given observed frequency, we 
show the closest mode of $\ell=3$ or 4: the frequency spectrum of these
modes is so dense that practically everywhere in the observed interval
a fitting within 0.1~c/d is possible. 

% *********************************** Table 4.
\begin{table*}
\begin{center}
\caption{Best fits for some FG Vir models}
% \label{fits}
%
\begin{tabular}{|ccc|crrc|crrc|crrc|crrc|}
\hline
\multicolumn{3}{|c|}{Observations}& \multicolumn{4}{c|}{Model 1}
& \multicolumn{4}{c|}{Model 3}& \multicolumn{4}{c|}{Model 8}
& \multicolumn{4}{c|}{Model 12}\\

N & $f \,$[c/d] &  id. & $\ell$ & $m$ & $f \,$[c/d] & $\mid$O-C$\mid$
& $\ell$ & $m$ & $f \,$[c/d] & $\mid$O-C$\mid$
& $\ell$ & $m$ & $f \,$[c/d] & $\mid$O-C$\mid$
& $\ell$ & $m$ & $f \,$[c/d] & $\mid$O-C$\mid$ \\
\hline
 1 & 12.716 &  1   & 1 & -1 & 12.821 & .105 & 1 & -1 & 12.714 & .002 & 1 & -1 & 12.814 & .098 %%@
& 1 & -1 & 12.719 & .003\\
   &        &      & 2 & -1 & 12.770 & .054 &   &    &        &      &   &    &        &      %%@
&   &    &        &     \\
\hline
 2 & 24.228 & 2/1  & 1 &  0 & 24.436 & .208 & 1 &  1 & 24.083 & .145 & 1 &  0 & 24.453 & .225 %%@
& 1 &  1 & 24.250 & .022\\
   &        &      & 3 & -2 & 24.331 & .103 & 3 & -2 & 24.128 & .100 & 3 & -2 & 24.381 & .153 %%@
& 3 & -2 & 24.308 & .080\\
\hline
 3 & 23.403 &  0   & 0 &  0 & 23.226 & .177 & 0 &  0 & 23.237 & .166 & 0 &  0 & 23.247 & .156 %%@
& 0 &  0 & 23.402 & .001\\
   &        &      & 3 &  1 & 23.285 & .118 &   &    &        &      & 3 &  1 & 23.317 & .086 %%@
&   &    &        &     \\
\hline
 4 & 21.052 &  2   & 1 & -1 & 20.828 & .224 & 1 & -1 & 20.758 & .294 & 1 & -1 & 20.667 & .385 %%@
& 1 & -1 & 20.739 & .313\\
   &        &      & 2 &  2 & 21.767 & .715 & 2 &  2 & 22.005 & .953 & 2 &  2 & 21.723 & .671 %%@
& 2 &  2 & 22.106 & .054\\
   &        &      & 3 & -2 & 21.079 & .027 & 3 & -3 & 21.092 & .040 & 3 & -2 & 20.961 & .091 %%@
& 3 & -3 & 21.024 & .028\\
   &        &      & 4 & -3 & 21.048 & .004 & 4 &  1 & 21.091 & .039 & 4 & -3 & 21.037 & .015 %%@
& 4 & -4 & 21.085 & .033\\
\hline
 5 & 19.868 &  2   & 2 & -2 & 19.996 & .128 & 2 & -2 & 19.800 & .068 & 1 &  1 & 19.893 & .025 %%@
& 2 & -2 & 19.774 & .094\\
   &        &      &   &    &        &      &   &    &        &      & 2 & -2 & 19.899 & .031 %%@
&   &    &        &     \\
\hline
 6 & 12.154 &  0   & 0 &  0 & 12.161 & .007 & 0 &  0 & 12.156 & .002 & 0 &  0 & 12.152 & .002 %%@
& 0 &  0 & 12.157 & .003\\
\hline
 7 &  9.656 & 1/2  & 1 & -1 &  9.609 & .047 & 2 & -2 &  9.705 & .049 & 1 & -1 &  9.250 & .406 %%@
& 2 &  2 &  9.656 & .000\\
   &        &      & 2 & -1 &  9.556 & .100 &   &    &        &      & 2 &  1 &  9.815 & .159 %%@
&   &    &        &     \\
   &        &      &   &    &        &      &   &    &        &      & 3 & -3 &  9.661 & .005 %%@
&   &    &        &     \\
\hline
 8 &  9.199 &  2   & 1 &  1 &  9.228 & .029 & 2 &  0 &  9.267 & .068 & 1 & -1 &  9.250 & .051 %%@
& 1 & -1 &  9.143 & .056\\
   &        &      & 2 &  0 &  9.242 & .043 &   &    &        &      & 2 & -2 &  8.940 & .259 %%@
&   &    &        &     \\
\hline
 9 & 19.228 & (0)  & 1 &  0 & 19.204 & .024 & 1 &  0 & 19.217 & .011 & 2 &  0 & 19.265 & .037 %%@
& 2 &  0 & 19.326 & .098\\
\hline
10 & 20.288 & (1)  & 1 &  1 & 20.162 & .126 & 1 &  1 & 20.293 & .005 & 1 &  0 & 20.385 & .097 %%@
& 1 &  1 & 20.206 & .082\\
   &        &      & 3 &  0 & 20.333 & .045 &   &    &        &      &   &    &        &      %%@
&   &    &        &     \\
\hline
11 & 24.200 &  -   & 1 &  0 & 24.436 & .236 & 1 &  1 & 24.083 & .117 & 1 &  0 & 24.453 & .253 %%@
& 1 &  1 & 24.250 & .050\\
   &        &      & 3 & -2 & 24.331 & .131 & 3 & -2 & 24.128 & .072 & 4 &  1 & 24.074 & .126 %%@
& 3 & -2 & 24.308 & .108\\
\hline
12 & 16.074 &  -   & 1 &  0 & 16.191 & .117 & 1 &  0 & 16.170 & .096 & 1 &  0 & 16.124 & .050 %%@
& 1 &  0 & 16.211 & .137\\
   &        &      & 2 &  2 & 15.946 & .128 & 3 & -1 & 16.079 & .005 & 3 & -1 & 16.016 & .058 %%@
& 2 &  2 & 16.123 & .049\\
   &        &      & 4 &  0 & 16.060 & .014 & 4 &  0 & 16.082 & .008 & 4 & -1 & 16.010 & .064 %%@
& 4 &  2 & 16.114 & .040\\
\hline
13 & 34.119 &  -   & 2 &  0 & 34.231 & .112 & 2 &  0 & 34.221 & .102 & 2 &  1 & 33.916 & .203 %%@
& 2 &  2 & 33.970 & .149\\
   &        &      &   &    &        &      &   &    &        &      & 3 & -2 & 34.172 & .053 %%@
& 3 &  0 & 34.019 & .100\\
\hline
14 & 21.232 &  -   & 1 & -1 & 20.828 & .404 & 1 & -1 & 20.758 & .474 & 1 & -1 & 20.667 & .565 %%@
& 1 & -1 & 20.739 & .493\\
   &        &      & 3 & -2 & 21.079 & .153 & 3 & -3 & 21.092 & .140 & 2 &  2 & 21.723 & .491 %%@
& 3 & -3 & 21.024 & .208\\
   &        &      & 4 & -4 & 21.368 & .136 & 4 &  0 & 21.367 & .135 & 4 &  0 & 21.167 & .065 %%@
& 4 &  0 & 21.147 & .085\\
\hline
15 & 11.110 &  -   & 2 & -1 & 10.975 & .135 & 2 & -2 & 11.099 & .011 & 2 &  2 & 11.423 & ..313 %%@
& 2 &  2 & 11.649 & .539\\
   &        &      & 3 &  2 & 11.099 & .011 & 3 & -2 & 11.074 & .036 & 3 &  0 & 10.942 & .168 %%@
& 3 & -1 & 11.024 & .086\\
   &        &      & 4 &  0 & 11.092 & .018 & 4 &  0 & 11.085 & .025 & 4 &  4 & 10.983 & .127 %%@
& 4 &  2 & 11.059 & .051\\
\hline
17 & 33.056 &  -   & 2 &  2 & 33.316 & .260 & 1 & -1 & 32.609 & .447 & 1 & -1 & 32.785 & .271 %%@
& 1 & -1 & 32.959 & .097\\
   &        &      & 4 & -1 & 33.020 & .036 & 4 & -2 & 33.152 & .096 & 3 &  1 & 33.092 & .036 %%@
& 4 &  0 & 33.032 & .024\\
\hline
18 & 21.551 &  -   & 2 &  2 & 21.767 & .216 & 2 &  2 & 22.005 & .454 & 2 &  2 & 21.723 & .172 %%@
& 2 &  2 & 22.106 & .555\\
   &        &      & 3 & -3 & 21.433 & .118 & 4 & -1 & 21.642 & .091 & 4 & -1 & 21.566 & .015 %%@
& 4 & -1 & 21.421 & .130\\
\hline
19 & 28.140 &  -   & 1 &  1 & 27.959 & .181 & 1 &  1 & 28.125 & .015 & 2 & -1 & 28.247 & .107 %%@
& 2 &  1 & 28.062 & .078\\
   &        &      & 4 &  1 & 28.205 & .065 & 3 & -2 & 28.133 & .007 & 3 & -1 & 28.101 & .039 %%@
& 3 & -1 & 28.163 & .023\\
\hline
20 & 11.195 &  -   & 2 & -2 & 11.277 & .082 & 2 & -2 & 11.099 & .096 & 2 &  2 & 11.423 & .228 %%@
& 2 &  2 & 11.649 & .454\\
   &        &      & 3 & -2 & 11.281 & .086 & 3 & -3 & 11.309 & .114 & 3 & -1 & 11.289 & .094 %%@
& 3 &  3 & 11.223 & .028\\
   &        &      & 4 &  2 & 11.231 & .036 & 4 &  3 & 11.181 & .014 & 4 & -2 & 11.285 & .090 %%@
& 4 &  4 & 11.175 & .020\\
\hline
21 & 24.354 &  -   & 1 &  0 & 24.436 & .082 & 1 &  0 & 24.413 & .059 & 1 &  0 & 24.453 & .099 %%@
& 1 &  1 & 24.250 & .104\\
   &        &      & 3 & -2 & 24.331 & .023 & 3 & -3 & 24.349 & .005 & 3 & -2 & 24.381 & .027 %%@
& 3 & -2 & 24.308 & .046\\
   &        &      & 4 &  0 & 24.398 & .044 &   &    &        &      & 4 &  0 & 24.457 & .103 %%@
& 4 &  1 & 24.339 & .015\\
\hline
22 & 11.870 &  -   & 2 &  1 & 12.029 & .159 & 2 &  2 & 11.859 & .011 & 2 &  1 & 11.818 & .052 %%@
& 2 &  1 & 11.915 & .045\\
   &        &      & 4 &  4 & 11.846 & .024 &   &    &        &      & 3 & -3 & 11.969 & .099 %%@
&   &    &        &     \\
\hline
\end{tabular}
\end{center}
\end{table*}

Model 12 seems to be the best-fitting model in our series. Note
the excellent agreements in both frequency and $\ell$--identifications for
all ten dominant frequencies. 
The mean difference (O-C) is about 0.04~c/d for these modes. The fit
for most of the remaining frequencies is also good, once $\ell=3$
$\ell=4$ modes are considered. Possible discrepances at low frequencies
do not appear serious to us due to the strong sensitivity of g-modes
to model parameters. Because of the effects of avoided crossing some
of the quadrupole modes of higher
order are also quite sensitive to small variations of parameters 
(cf. $\ell=2$ modes for models 1 and 2 at approximately 32~c/d).

In Fig.~8 we compare the range of observed frequencies and of excited modes 
explicitly for the reference model as well as three other models. 
FG~Vir is located in the HR-diagram in the middle of the instability strip
(see, for example, Breger \& Pamyatnykh 1998). Consequently, the driving of
oscillations is effective in a wide frequency region which extends over 7
radial overtones. 
The independence of the driving efficiency on the mode degree, $\ell$, is
typical for the oscillations excited by the opacity mechanism. 

We also note that the observed frequency spectrum is divided clearly into two groups:
at 9-13 c/d and at 19-25 c/d. However, the present models do not predict any
instability gap in frequency, as can be seen in Fig.~8.

\begin{figure}
\centering
\includegraphics*[width=88mm]{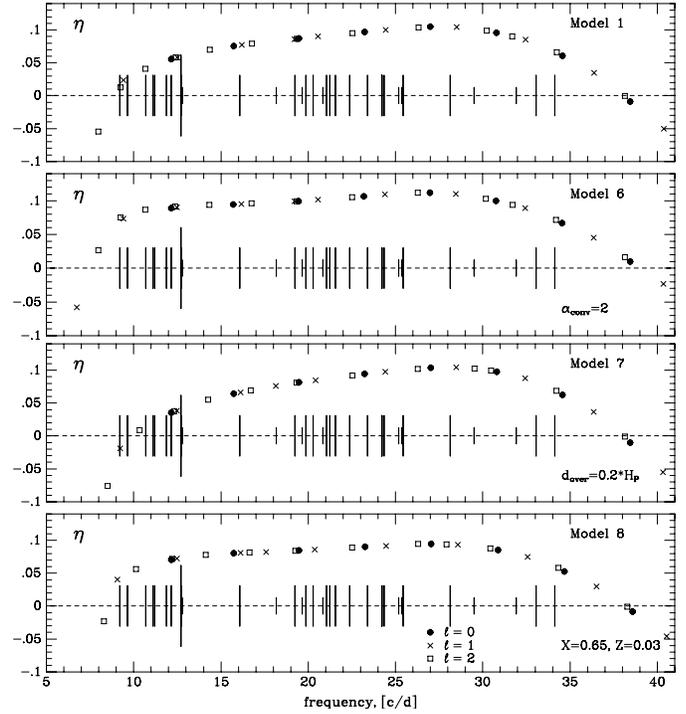}
\caption{
Normalized growth-rates, $\eta$,  of low-degree oscillation modes of some
FG~Vir models from Table 4. Only axisymmetric modes
($m=0$) are shown. Positive values correspond to unstable modes.
Vertical lines mark observed frequencies from Table~1, probable frequencies
$f_{25}$ to $f_{32}$ are shorter. 
The longest line at 12.716~c/d corresponds to the mode $f_1$
with the highest amplitude}
\end{figure}

\subsection{Rotational mode coupling}

An additional factor which influences the frequencies of a rotating star
is the coupling between modes with close frequencies whose azimuthal orders, 
$m$, are the same and whose degrees, $\ell$, are also the same or differ by 2.
The effect was described and discussed in detail
by Dziembowski \& Goode (1992) and by Soufi et al. (1998). 
The frequency distance between two near-degenerate modes increases when
coupling is taken into account.
The significance of this rotational frequency perturbation was demonstrated
by Pamyatnykh et al. (1998) in application to XX~Pyx. It was shown that
at rotational velocity of about 90 km/s the frequency shifts of coupled
radial and quadrupole (or dipole and octupole) overtones achieve
0.1--0.2 c/d. 
Therefore, in particular, a significant change 
of radial frequency ratios may be expected. 

We estimated this effect in some of our FG~Vir models and
found it to be unimportant at rotational velocities of about and less
than 45 km/s. For example, for the reference model 1, the frequencies of
the radial fundamental mode, $f(p_1)$, and of the third overtone, $f(p_4)$,
are changed due to coupling with closest axisymmetric quadrupole modes
by -0.0035 c/d and 0.0091 c/d, respectively. 
The effect is much stronger
for more rapidly rotating models: for model 5 with $V_{\rm rot}$ = 90~km/s,
the radial fundamental mode and third overtone are shifted by -0.0436 c/d
and 0.1545 c/d, respectively, which results in the change of the
frequency ratio from 0.5266 to 0.5212. 
As another example, we were able to reproduce the observed
ratio $f_6 / f_3 = 0.5193$ as the radial frequency ratio for a model with
initial abundances $X=0.65$, $Z=0.03$ and with $V_{\rm rot}$ = 91-92~km/s.
However, the rapid rotation seems to be rather improbable for FG~Vir with
$v \sin i = 21 \pm 1\, $km/s, as it was discussed by Viskum et al. (1998).

Note that the coupling effect is
higher for higher overtones: for model 5, the shift of the radial 
sixth overtone frequency (34.306 c/d) is 0.520 c/d due to interaction
with the closest axisymmetric quadrupole mode of 34.219 c/d which is shifted
by the same quantity 0.520 c/d in the opposite direction. Moreover,
$\ell = 2$ modes are affected by coupling with $\ell = 4$ modes, and
so on. We conclude that for rapidly rotating models it is necessary to
take the rotational coupling into account in attempts to fit the observed
frequency spectrum with the theoretical one.

The rotational coupling results also in a mutual contamination of amplitudes
of spherical harmonic components of interacting modes (Soufi et al 1998,
see examples in Table 4 of Pamyatnykh et al. 1998).
This adversely affects the mode discrimination
by means of multicolor photometry (see the footnote in section 3) and should
influence spectroscopic determinations as well. However, 
this effect was found to be important only in models with more rapid
rotation than found for FG~Vir. A more detailed discussion of the rotational
mode coupling problem in connection with the interpretation of the
observed multifrequency spectrum will be given by Dziembowski \& Goupil (1998).

\bigskip
\acknowledgements
%________________________________________ Do not leave a blank line here !
We are grateful to M.~Viskum and S.~Frandsen for making the photometry
used in their FG Vir paper available to us and W.~Dziembowski for
stimulating discussions. Part of the investigation has been supported by the
Austrian Fonds zur F\"{o}rderung der wissenschaftlichen Forschung,
project number S7304. AAP acknowledges partial financial support
by the Polish Committee for Scientific Research (grant 2-P03D-014-14)
and by the Russian Foundation for Basic Research (grant 98-02-16734).

\end{document}